\documentclass{article}

\usepackage[utf8]{inputenc}

\usepackage{authblk}
\usepackage{amsmath}
\usepackage{natbib}
\usepackage{color,soul}

\usepackage{geometry}
\usepackage{graphicx}

\usepackage{caption}

\usepackage{subcaption}

\title{Combining Accelerometer and Gyroscope Data in Smartphone-Based Activity Recognition using Movelets}
\author[1]{Emily J. Huang}
\author[2]{Kebin Yan}
\author[3]{Jukka-Pekka Onnela}
\affil[1]{Department of Mathematics and Statistics, Wake Forest University}
\affil[2]{Department of Biostatistics, Epidemiology, and Informatics, University of Pennsylvania}
\affil[3]{Department of Biostatistics, Harvard University}

\date{}

\begin{document}

\maketitle

\indent Correspondence: huange@wfu.edu

\abstract{Physical activity patterns can be informative about a patient's health status. Traditionally, activity data have been gathered using patient self-report. However, these subjective data can suffer from bias and are difficult to collect over long time periods. Smartphones offer an opportunity to address these challenges. The smartphone has built-in sensors that can be programmed to collect data objectively, unobtrusively, and continuously. Due to their widespread adoption, smartphones are also accessible to most of the population. A main challenge in smartphone-based activity recognition is extracting information optimally from multiple sensors to identify the unique features of different activities. In our study, we analyze data collected by the accelerometer and gyroscope, which measure the phone's acceleration and angular velocity, respectively. We propose an extension to the ``movelet method'' that jointly incorporates both sensors. We also apply this joint-sensor method to a data set we collected previously. The findings show that combining data from the two sensors can result in more accurate activity recognition than using each sensor alone. For example, the joint-sensor method reduces errors of the gyroscope-only method in differentiating between standing and sitting. It also reduces errors of the accelerometer-only method in classifying vigorous activities.}

\section{Introduction}

A person's physical activity pattern is related to her or his health status and thus provides an overall health profile. For example, people recovering from a major surgery may move less or may move more slowly than what is typical for them, and the duration of changed activity pattern can inform about their recovery trajectory \citep{panda2020}. Similarly, an increase in purposeless movement (e.g., pacing and inability to sit still) may be a symptom of depression \citep{dsm2013}. Thus, it may be beneficial to monitor the relevant daily activities of people at risk of developing a health condition. This kind of information has traditionally been gathered by having patients take surveys, provide interviews, or complete diaries \citep{sylvia2014}. Although these patients' self-reports, as firsthand accounts, are beneficial, they also have some limitations \citep{sallis2000}. For example, the self-report data are often questioned for their natural proclivity to bias: patients may downplay certain tendencies because they like to be viewed as ``normal.'' Self-report may overestimate exercise levels for ``good social image'' \citep{adams2005}. Patients can also provide inaccurate reports unintentionally because human memory is prone to mistakes \citep{shephard2003}. 

Researchers have sought to find new objective ways for collecting more reliable physical activity data to  complement the self-reported data. The rise in smartphone adoption and usage offers a unique opportunity to revolutionize patient health status monitoring in both research settings and in clinical practice. The built-in smartphone sensors, such as the GPS, accelerometer, gyroscope, and magnetometer, can track location and movement continuously and unobtrusively.  These \emph{in situ} data can be collected to quantify daily activity objectively. Smartphone data collection does not require outfitting patients with additional instruments and thus can be done over long periods of time \citep{onnela2021}. Smartphones are also widely accessible to the population. Based on surveys by the Pew Research Center, about 85\% of U.S. adults own smartphones as of 2021, which is almost 2.5 times from 10 years ago \citep{pew2021}. 
The field of digital phenotyping emerged to take advantage of this new technological breakthrough and the vast amount of smartphone sensor data. Digital phenotyping is defined as the ``moment-by-moment quantification of the individual-level human phenotype \emph{in situ} using data from smartphones and other personal digital devices'' \citep{torous2016}. This approach uses smartphones to capture high-throughput data to learn about cognitive, behavioral, and social phenotypes in free-living settings. 

Human activity recognition (HAR) using smartphones has proliferated in recent years \citep{straczkiewicz2021}. The two components of HAR are data collection and data analysis.  The data collection setting requires careful thought about various questions, such as choosing the appropriate sensors, sampling frequency, study environment, and smartphone placement. Some studies use a single sensor only whereas others utilize multiple sensors simultaneously \citep{shoaib2014, gu2015, capela2016, hnoohom2017, filntisis2020}. In our study, we focus on data collected from both the accelerometer and gyroscope sensors in the smartphone. An accelerometer measures the acceleration of a phone along each of three orthogonal axes of a Cartesian coordinate system. The $x$-axis and $y$-axis are in the plane of the phone's screen, with $x$ pointing right and $y$ pointing to the top of the phone. The $z$-axis points up through the phone, following the right hand rule. A gyroscope measures the angular velocity of a phone about three orthogonal axes. In previous HAR studies, a variety of sampling frequencies (samples per second) have been used (e.g., 1 Hz or even 100 Hz), commonly ranging between 20 to 30 Hz \citep{straczkiewicz2021}.  In our study, we sampled accelerometer and gyroscope data at a frequency of 10 Hz, which is sufficient for capturing most daily activities.

With improvements in technology, cost, and quality of data collection, the main challenge in HAR is shifting to data analysis, i.e., to extract the activities from the sensor data accurately and robustly \citep{trifan2019,onnela2021, straczkiewicz2021}. In general, a given data analysis procedure can be divided into three steps: preprocessing, feature extraction, and activity classification \citep{straczkiewicz2021}. Preprocessing prepares the data for the analysis at hand. For example, it might include removal of irrelevant high-frequency fluctuations (noise). The feature extraction step involves selecting and extracting representative features from the data. In activity classification, the extracted features are first associated with physical states or physical activities using statistical models. These models are then used to classify activities for new data.

In this paper, our main focus is feature extraction and activity classification. Previous HAR studies have used a variety of feature extraction and activity classification techniques. A rapidly developing field is the application of deep learning, which automates both feature extraction and activity classification. Using multiple layers in the network, the deep learning procedure identifies optimal features from the raw data itself, without any human intervention \citep{demrozi2020}. Some studies show that this approach can yield highly accurate results in activity classification \citep{ha2016, dhanraj2019}. However, there are limitations and challenges in application. First, a vast amount of data are required to train a deep learning algorithm. Second, the model is usually used as a black box, and the extracted features from the multi-layered procedure can be difficult to interpret \citep{demrozi2020}. These result in difficulties in algorithm improvement.

A more traditional approach of data analysis is to view the data in short segments, referred to as windows. This approach allows us to examine the data directly and choose which features to extract through the most appropriate methods. Subsequently, a model may be constructed from training data to connect the selected features to activities. In this paper, we adopted the ``movelet method'' for feature extraction and activity classification,  which was developed by \citet{bai2012} and later augmented by \citet{huang2020}.  The unique aspect of this method is that it is tailored to each individual patient by constructing a personal dictionary of windows for different types of activities from her/his training data.  
The patient's activities are inferred by comparing new data with the data in the dictionary. Given that the movelet method makes activity classifications for each person based on her/his own dictionary, the classifications are personalized to the individual's unique data patterns, and therefore account for factors such as the person's height, weight, age, and health condition. Models built using training data from one cohort (e.g., young, healthy people) may perform poorly when applied to another group (e.g., older adults or patients with illnesses) \citep{albert2012,del2014}. Compared to deep learning, the movelet method only requires a small amount of training data (a few seconds per activity) to build the person's dictionary. The movelet method uses pattern recognition or pattern matching, which is different from methods that use summary statistics of the patterns (e.g., mean, variance, interquartile range) within each window as features. More sophisticated machine learning methods can be applied to link those summary statistics to activities, but not without requiring substantially more training data. Common models include $k$-nearest neighbors, naive Bayes, support vector machines, decision trees, and random forests. 

Some previous studies have used the movelet method to classify activities with a single sensor \citep{bai2012, huang2020}. \citet{bai2012} analyzed data collected by a body-worn accelerometer. \citet{huang2020} applied the method to smartphone accelerometer data and separately to smartphone gyroscope data. They also proposed a method to quantify the uncertainty in the classifications. The results showed that the smartphone accelerometer and gyroscope each had strengths in picking up different activities. For example, the accelerometer outperformed gyroscope in accurately classifying standing and sitting, while the gyroscope outperformed the accelerometer in accurately classifying walking. In this study, we analyzed smartphone accelerometer data and gyroscope data jointly. Our hypothesis is that combined information from both acceleration and angular velocity would improve the accuracy of classification because the individual sensors capture different aspects of movement.  The previous study by \citet{he2014} used multiple accelerometers fixed to different parts of the body. They found improvements in classification accuracy using the integrated information from the multiple instruments. Although the smartphone is different from body-worn instruments, we expected its multiple sensors to provide similar benefit in improving classification accuracy. In comparison to multiple body-worn instruments, the smartphone has the advantage that it is compact, convenient to carry, and can be used over long time periods. In this paper, we report the results from our extended version of the original movelet method that jointly incorporates smartphone accelerometer and gyroscope data. Our R code is provided in the Supplement.

The paper is organized as follows. Section \ref{s:methods} describes our data set and presents the method for incorporating accelerometer and gyroscope data in the movelet method. In Section \ref{s:results}, we present the results of applying this method to the study data set. We also compare the results to those from applying the movelet method to accelerometer data only and to gyroscope data only. Section \ref{s:discussion} summarizes the results and discusses potential areas of future research.

\section{Materials and Methods}
\label{s:methods}

\subsection{Study Data Set}
\label{s:study_description}

The data set used in this paper is from a study we conducted in 2018. The study included two female and two male participants, ranging in age from 27 to 54. Characteristics of the participants are provided in Table S1 of \citet{huang2020}, including sex, height, weight, and dominant hand. For full disclosure, Participant 1 is an author of this paper. Each participant had a study visit in which she/he performed a series of activities while wearing a study iPhone in the front right pants pocket and another study iPhone in the back right pants pocket. Throughout this paper, we focus on the front pocket phone, and refer to it as the phone. For each participant, we recorded the accelerometer and gyroscope measurements from the phone at a frequency of 10 Hz, i.e. 10 samples per second. Each participant was filmed throughout the experiment, and the video footage was used to manually annotate the sensor data with ground truth activity labels. The accelerometer and gyroscope data from the study with annotated activity labels are publicly available on Zenodo  \citep{emily_huang_2020_3925679}.

The data collection was divided into collection of training data and test data. During the training data collection, we collected brief snapshots of phone data as the participant performed designated activities. The activities included walking, standing, ascending stairs, descending stairs, sitting, and transitioning from sitting to standing (sit-to-stand), and transitioning from standing to sitting (stand-to-sit). For each participant, the duration of training data used in our analysis was 5 seconds per activity, except for sit-to-stand and stand-to-sit which are momentary transitions. The test data collection included six steps, where the participant followed a prescribed course of activities on the Harvard Longwood campus. For example, the course in Step 1 included walking, ascending stairs, standing, and descending stairs. The participant walked at different speeds in Step 3, and ascended and descended a long staircase in Step 6. In this paper, we use the training data, and the test data from Steps 1, 2, 3, 5, and 6. The test data from Step 4 is not analyzed in this paper. During Step 4, the participant repeated the same course four times with the phone reoriented in a different position each time. We discuss the issue of how the phone is carried in Section \ref{s:discussion}. Further description of the training and test data collection is provided in \citet{huang2020}.

\subsection{The Movelet Method: Single Sensor}

The movelet method, proposed by \citet{bai2012}, was originally designed for activity recognition from a body-worn accelerometer but can be applied to any single sensor. In our previous paper, we applied the movelet method separately to smartphone accelerometer data and smartphone gyroscope data \citep{huang2020}. The essence of the movelet method is to tailor its activity classifications to the patient's personal activity patterns. The method uses the following procedure.

First, the training data is  collected by having the participant perform a comprehensive list of daily life activities while collecting data from the sensor of interest (e.g., a smartphone accelerometer). Only a few seconds of training data are required per activity. In our analysis, we used up to five seconds of training data per activity. The training data are then used to build the ``dictionary'' for the participant. The entries in the dictionary are the different activities (e.g., walk, sit, stand) specified in the aforementioned list. Each activity entry consists of multiple movelets, where any given movelet is a one-second window of the sensor's tri-axial ($x$,$y$,$z$) data. The movelets for a given activity are obtained from the corresponding 5-second training data. For example, for the walk entry, one obtains the 1-second movelets by sliding a 1-second window one sample (0.1 seconds, the reciprocal of sampling frequency) at a time along the tri-axial data, until the right end of the one-second window meets the last point of the 5-second time series. The resulting number of movelets is 41 for the 5-second time series. The number of dictionary movelets of an entry depends on the data collection frequency and the duration of the training data.

Next, we perform activity classifications on new data (termed test data here) using the dictionary. For the test data, we also construct movelets by sliding a one-second window one sample at a time along the test data tri-axial time series. Each test movelet is then matched with one of the dictionary movelets based on the smallest discrepancy. The discrepancy metric uses Euclidean distance and is defined in Section \ref{s:discrepancy}. The activity label of the dictionary movelet is assigned to the test movelet as the classification. To make a classification for a given time point, we take a majority vote among the test movelets beginning at the time point and the subsequent nine following it. The activity that receives the most votes is taken as the classification for that time point. The rationale behind the majority vote process is that later movelets also contain activity information for the time point because human activities are continuous.

\subsection{The Movelet Method: Joint Sensors}
\label{s:movelet_joint}

In this paper, we propose an extension of the movelet method in which we use gyroscope and accelerometer data simultaneously. The motivation behind this is that we hypothesized that acceleration and angular velocity provide different physical information, and combining both could improve the accuracy of the activity classifications. The method follows the same procedure as above, except the one-second movelets include all six dimensions of data ($x$, $y$, $z$ from the accelerometer and gyroscope sensors) rather than only three dimensions (from a single sensor). Related work includes \citet{he2014} who applied the movelet method with data from multiple accelerometers fixed to different parts of the body. 

Although the accelerometer and gyroscope each followed a 10 Hz data collection schedule, they were not synchronized. Thus, for the application of the movelet method jointly to accelerometer and gyroscope data, we linearly interpolated the gyroscope data to the timestamps in the accelerometer data. This was done for the training data as well as the test data, so that dictionary and test movelets all had 6 measurements corresponding to a specific timestamp. We chose linear interpolation since it is simple and has been recommended for this type of data \citep{derawi2013}.

\subsection{Discrepancy Metric}
\label{s:discrepancy}

A discrepancy metric based on Euclidean distance was used to compare each test movelet to each dictionary movelet. The discrepancy is defined as follows. Consider any two movelets $M$ and $M'$. For the movelet $M$, we define $\boldsymbol{x}$ as the vector of time series data of the $x$-axis for one smartphone sensor (accelerometer or gyroscope) in a 1-second window. The vectors $\boldsymbol{y}$ and $\boldsymbol{z}$ are defined analogously. To be precise, we have $\boldsymbol{x}=\left(x_1,x_2,\ldots, x_n\right)$, $\boldsymbol{y}=\left(y_1,y_2,\ldots, y_n\right)$, and $\boldsymbol{z}=\left(z_1,z_2,\ldots, z_n\right)$, where $n = 10$ since our data sampling frequency is 10 Hz. For another movelet $M'$, we similarly define $\boldsymbol{x'}=\left({x}'_1,{x}'_2,\ldots,{x}'_n\right)$, $\boldsymbol{y'}=\left({y}'_1,{y}'_2,\ldots,{y}'_n\right)$, and  $\boldsymbol{z'}=\left({z}'_1,{z}'_2,\ldots,{z}'_n\right)$. Using $\boldsymbol{a}$ and $\boldsymbol{a'}$ to represent a pair of vectors of a given dimension from $M$ and $M'$ respectively, the Euclidean distance for the dimension is defined as 
$$d_{L_2}\left(\boldsymbol{a},\boldsymbol{a}'\right)=\sqrt{\sum_{i=1}^n\left(a_i-a_i'\right)^2}.$$
For the single-sensor movelet method, we compute the distance for the $x$, $y$, and $z$ axes of the given sensor and average these three distances together. Thus, the discrepancy between the two movelets $M$ and $M'$ is:
$$\frac{1}{3}\left[d_{L_2}(\boldsymbol{x},
\boldsymbol{x'}) + d_{L_2}(\boldsymbol{y},\boldsymbol{y'})+ d_{L_2}(\boldsymbol{z},\boldsymbol{z'})\right].$$
For the joint-sensor movelet method, we compute the distance for the $x$, $y$, and $z$ axes of both the accelerometer and gyroscope and average these six distances together. The discrepancy between the two movelets $M$ and $M'$ is:
$$\frac{1}{6}\left[d_{L_2}(\boldsymbol{x_a},\boldsymbol{x_a'}) + d_{L_2}(\boldsymbol{y_a},\boldsymbol{y_a'})+ d_{L_2}(\boldsymbol{z_a},\boldsymbol{z_a'})+d_{L_2}(\boldsymbol{x_g},\boldsymbol{x_g'}) + d_{L_2}(\boldsymbol{y_g},\boldsymbol{y_g'})+ d_{L_2}(\boldsymbol{z_g},\boldsymbol{z_g'})\right],$$
where the $a$ and $g$ subscripts correspond to the accelerometer and gyroscope, respectively. As described in Section \ref{s:movelet_joint}, the vectors for the gyroscope in the joint-sensor analyses (i.e., $\boldsymbol{x_g}$, $\boldsymbol{y_g}$, and $\boldsymbol{z_g}$ for movelet $M$ and $\boldsymbol{x_g'}$, $\boldsymbol{y_g'}$, and $\boldsymbol{z_g'
}$ for movelet $M'$) are obtained by interpolating the original gyroscope data to the accelerometer timestamps, so that the two data sources are synchronized.

\subsection{Analysis Procedure}

We applied the extended version of the movelet method using accelerometer and gyroscope data jointly, and we also compared its classification accuracy to the original (i.e., single-sensor) movelet method using the accelerometer data only and the gyroscope data only. Table \ref{table:analysis_procedure} summarizes the key points of the analysis procedure. In the gyroscope-only analyses, we used the original gyroscope data rather than interpolated gyroscope data. This was done to mimic how an analysis using only gyroscope data  would be performed in practice. Applying the movelet method to the original gyroscope data resulted in an activity classification for each gyroscope timestamp. Since the accelerometer-only and joint-sensor analyses yielded classifications at the accelerometer timestamps, we then computed activity classifications for each accelerometer timestamp by taking the classification for the closest gyroscope timestamp. The R code for this paper is provided in the Supplement.

\begin{table}[h]
\centering
\begin{tabular}{|c|c|c|} 
 \hline
 \textbf{Sensor} & \textbf{Training Data} & \textbf{Test Data} \\
 \hline
  & & \\
  & Accelerometer: 5 seconds of the & Accelerometer: \\
 &  original training data for each activity & Original test data \\ 
 \textbf{Joint-sensor} &  & \\
 & Gyroscope: 5 seconds of the training  &  Gyroscope: \\
 & data for each activity, interpolated & Test data interpolated to \\
 & to the accelerometer timestamps & the accelerometer timestamps\\
  & & \\
 \hline
  & & \\
 \textbf{Accelerometer only} & 5 seconds of the original training & Original test data \\
 & data for each activity & \\ 
  & & \\
 \hline
 & &  \\
 \textbf{Gyroscope only} & 5 seconds of the original training &  Original test data\\
 & data for each activity &   \\ 
 & & \\
 \hline
 
 \hline
\end{tabular}
\caption{Analysis Procedure. \textit{This table summarizes the analysis procedure of this paper. The rows of the table show the methods, including the joint-sensor method, the accelerometer-only method, and the gyroscope-only method. For each method, the ``Training Data'' column indicates the amount of training data used per activity, and whether linear interpolation was applied to the data. The ``Test Data'' column indicates whether linear interpolation was applied to the test data.}}
\label{table:analysis_procedure}
\end{table}

\section{Results}
\label{s:results}

\subsection{Training Data Example}

We built a separate dictionary for each participant using her/his training data, with up to 5 seconds per activity. As an example, Figure 1 shows the training data of Participant 1. For both sensors, the signals for standing and sitting are flat lines, while those for walking, ascending stairs, and descending stairs are variable and quasi-periodic. The signals for the sit-to-stand and stand-to-sit activities are smooth curves that occur over a brief period of time. 

In the accelerometer training data, we observe that each activity has its own characteristic signature. For example, standing and sitting can be distinguished since the $x$, $y$, and $z$ data are at different levels. In standing, $y$ is approximately +0.98g and $x$ and $z$  are close to -0.15g and 0.02g, respectively, likely corresponding to the phone being positioned vertically. In sitting, $x$ is approximately -0.73g, $z$ is approximately 0.64g and $y$ is approximately 0.26g. This may correspond to the phone being in the right pocket. The curves for stand-to-sit are a smooth transition from the coordinates for standing to sitting, and vice versa for sit-to-stand.  Also, the patterns for walking, ascending stairs, and descending stairs differ from each other (e.g., the shapes in the $z$-axis data are different). Moreover, within an activity, the three axes also show different signatures (e.g., different fluctuations and peaks). For example, in walking, the $x$ and $y$ have an out-of-phase tendency in their extremes but the $z$ shows more degrees of freedom.

In the gyroscope training data, we also observe characteristic signatures for different activities. These signatures are different from what we saw in the accelerometer data. In both standing and sitting, the $x$, $y$, and $z$ are all approximately 0 radians/second because the phone is not rotating. Thus, given a short segment of gyroscope signal, we may not be able to distinguish sitting from standing. However, the transitions sit-to-stand and stand-to-sit can be distinguished by gyroscope. The sit-to-stand transition shows smooth arcs that start and end at 0. In contrast, the stand-to-sit transition is more variable in the beginning and then approaches 0 smoothly. For the activities of walking, ascending stairs, and descending stairs, the patterns appear to be more periodic for the gyroscope than for the accelerometer. The three axes are also more synchronized. Therefore, the accelerometer and gyroscope data are complementary to each other.

\begin{figure}[h!]
\centering
\includegraphics[width=\linewidth]{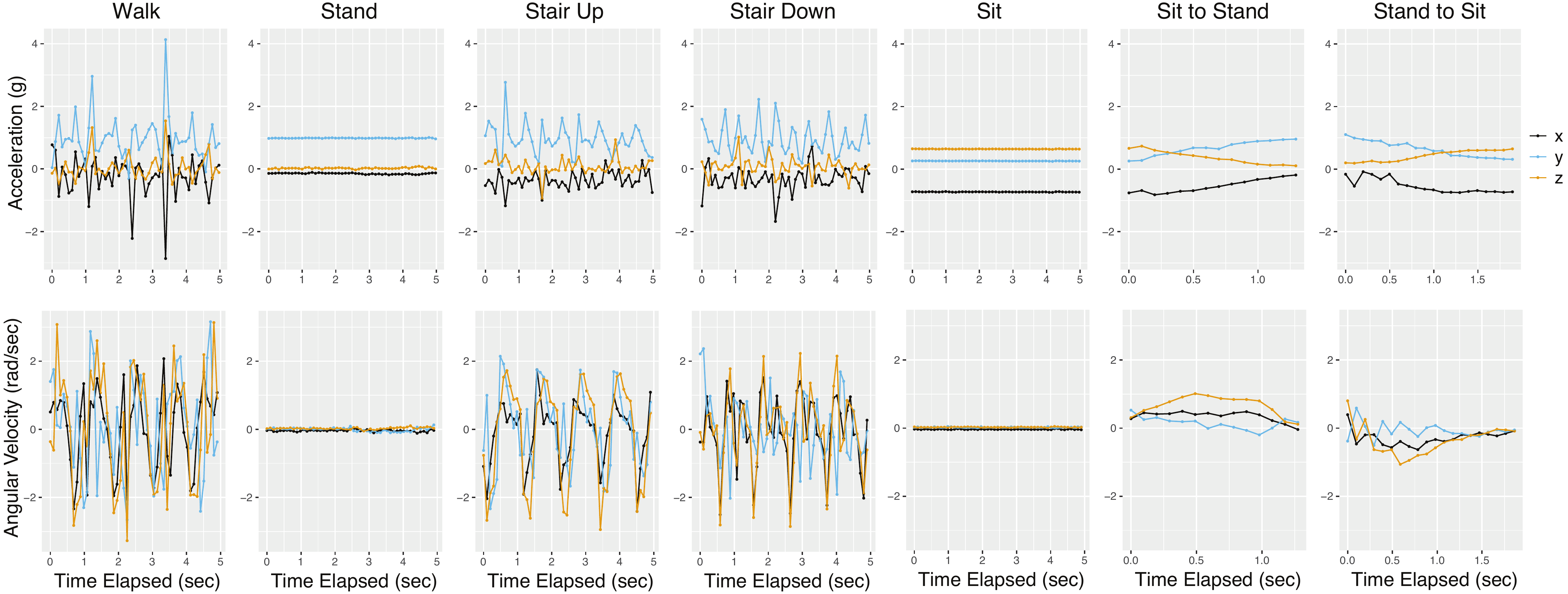}
\caption{Training Data for Participant 1. \textit{This figure shows the training data for Participant 1. The accelerometer training data is presented in the first row, and the gyroscope training data in the second row. The accelerometer measures the acceleration of the smartphone along the $x$, $y$, and $z$ axes. The acceleration data are measured in units of $g$ (i.e., $9.81$ $m/s^2$). The gyroscope measures the angular velocity of the smartphone projected onto the $x$, $y$, and $z$ axes. The angular velocity data are measured in units of radians per second (rad/sec). For both types of data, the $x$-axis is shown in black, the $y$-axis in blue, and the $z$-axis in orange.}}
\label{fig:P1_Training}
\end{figure}

\subsection{Results for Participant 3}
\label{s:results_participant3}
 
We applied the movelet method to Steps 1, 2, 3, 5, and 6 of each participant's test data. Table \ref{table:testData} shows the amount of test data per participant. In this section, we present the joint-sensor results of Participant 3 as an example. We also compare them to the accelerometer-only and gyroscope-only results. In Section \ref{s:results_allParticipants}, we then summarize the joint-sensor results of all participants. 

\begin{table}[h]
\centering
\begin{tabular}{c|c|c|c|c} 
    \hline
    & Participant 1 & Participant 2 & Participant 3 & Participant 4\\
    \hline \hline
    walk & 3288 & 3135 & 3287 & 3246 \\ 
    stand & 201 & 150 & 240 & 203 \\ 
    stair up & 321 & 302 & 361 & 342 \\ 
    stair down & 404 & 274 & 324 & 302 \\ 
    sit & 291 & 232 & 242 & 251 \\ 
    sit-to-stand & 21 & 20 & 20 & 20 \\ 
    stand-to-sit & 20 & 20 & 20 & 20 \\ 
    revolving door & 50 & 31 & 20 & 30 \\
    \hline
\end{tabular}
\caption{Amount of Test Data per Participant. \textit{This table shows the total number of samples per participant across Steps 1, 2, 3, 5, and 6 of the test data collection. The sampling frequency was approximately 10 Hz (i.e., 10 samples per second). For each participant, we present the number of samples of each activity type. The activity types observed in the test data collection included walking (``walk''), standing (``stand''), ascending stairs (``stair up''), descending stairs (``stair down''), sitting (``sit''), the sit-to-stand transition (``sit-to-stand''), the stand-to-sit transition (``stand-to-sit''), and going through a revolving door (``revolving door''). The ``walk'' activity includes slow, normal, and fast paced walking.}}
\label{table:testData}
\end{table}

\begin{figure}[h!]
    \includegraphics[width=\linewidth]{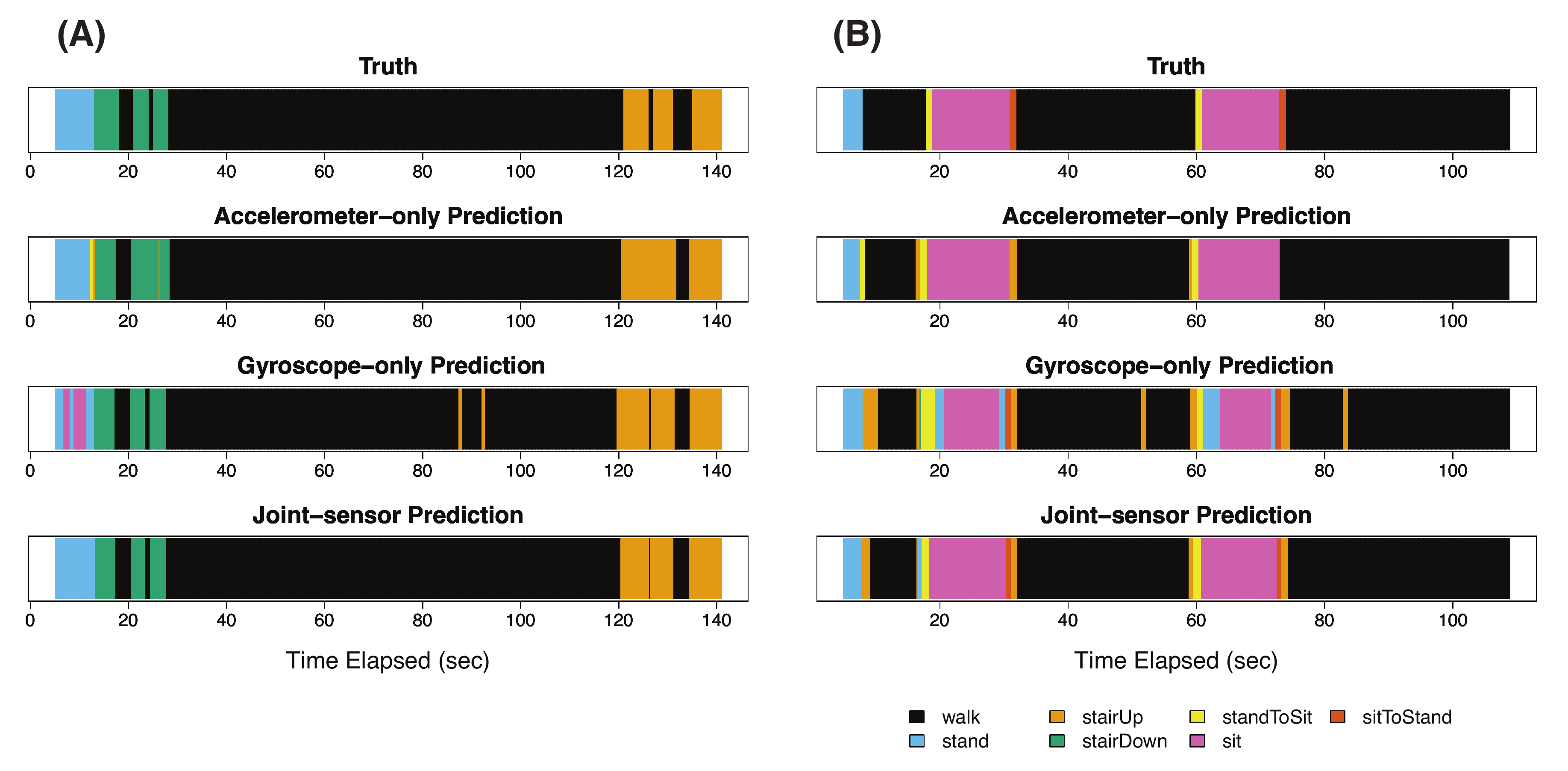}
    \caption{Steps 1 and 2 for Participant 3. \textit{Panels A and B show the results for Participant 3 in Steps 1 and 2, respectively. For each panel, four figures are shown. The top figure gives the true activity labels based on video footage. The second figure shows the activity classifications from the accelerometer-only method, the third figure shows the classifications from the gyroscope-only method, and the fourth shows the classifications from the joint-sensor method. In each row, the horizontal axis gives the time elapsed, measured in seconds.}}
    \label{fig:P3_S1_S2_CompGraph}
\end{figure}

Figure \ref{fig:P3_S1_S2_CompGraph}A shows the results for Participant 3 in Step 1. In Step 1, the participant performed the activities of standing (blue), descending stairs (green), walking (black), and ascending stairs (orange). The top image in Figure \ref{fig:P3_S1_S2_CompGraph}A shows the true activity labels based on video footage, the second image shows the classifications for accelerometer-only, the third shows the classifications for gyroscope-only, and the fourth shows the classifications for the joint-sensor method. 

Both the accelerometer-only and gyroscope-only methods recognized descending stairs and ascending stairs accurately. The accelerometer-only method also correctly recognized standing. However, the gyroscope-only method misclassified 47.5\% of the standing period as sitting (pink). This is consistent with the visual examination of the training data that sitting and standing periods are not distinguishable from gyroscope data. For the walking (black) period, the accelerometer-only method matched the truth accurately. While the gyroscope-only classifications were also mostly accurate, there were two segments that were characterized as ascending stairs. Whether this reflects small changes on the ground or are an artifact needs further examination. 

Combining the accelerometer and gyroscope data, the accurate ascending stair and descending stair classifications were preserved. Moreover, standing was classified correctly, improved from the gyroscope-only classifications. The walking classifications were also accurate. In particular, the joint-sensor method corrected the two segments misclassified as ascending stairs from gyroscope alone. The joint-sensor method also captured the brief periods of walking that fell between ascending or descending stairs, which were sometimes smoothed over by the accelerometer alone.

Figure \ref{fig:P3_S1_S2_CompGraph}B shows the results for Step 2. During this step, the participant performed the activities of standing (blue), walking (black), stand-to-sit (yellow), sitting (pink), and sit-to-stand (red). The joint-sensor method and accelerometer-only method classified the sitting periods (pink) accurately. The gyroscope-only method characterized these periods correctly most of the time, but misclassified the start and end of each sitting period as standing. The walking periods were classified well by all three methods, with the joint-sensor and accelerometer-only performing the best. Like in Step 1, the joint-sensor method had improved accuracy for walking compared to the gyroscope-only method, correcting the two false positives of ascending stairs that occurred during the second and third walking periods. Compared to gyroscope alone, the joint-sensor method also reduced the length of the ascending stair error at the beginning of the first walking period. All three methods picked up the two stand-to-sit transitions (yellow). However, the gyroscope-only method overestimated the length of the first stand-to-sit transition. The joint-sensor method reduced the length closer to the truth. The sit-to-stand transitions (red) were almost entirely missed by the accelerometer-only method, but they were captured by the joint-sensor and gyroscope-only methods. 

\begin{figure}[h!]
    \centering
    \includegraphics[width=\linewidth]{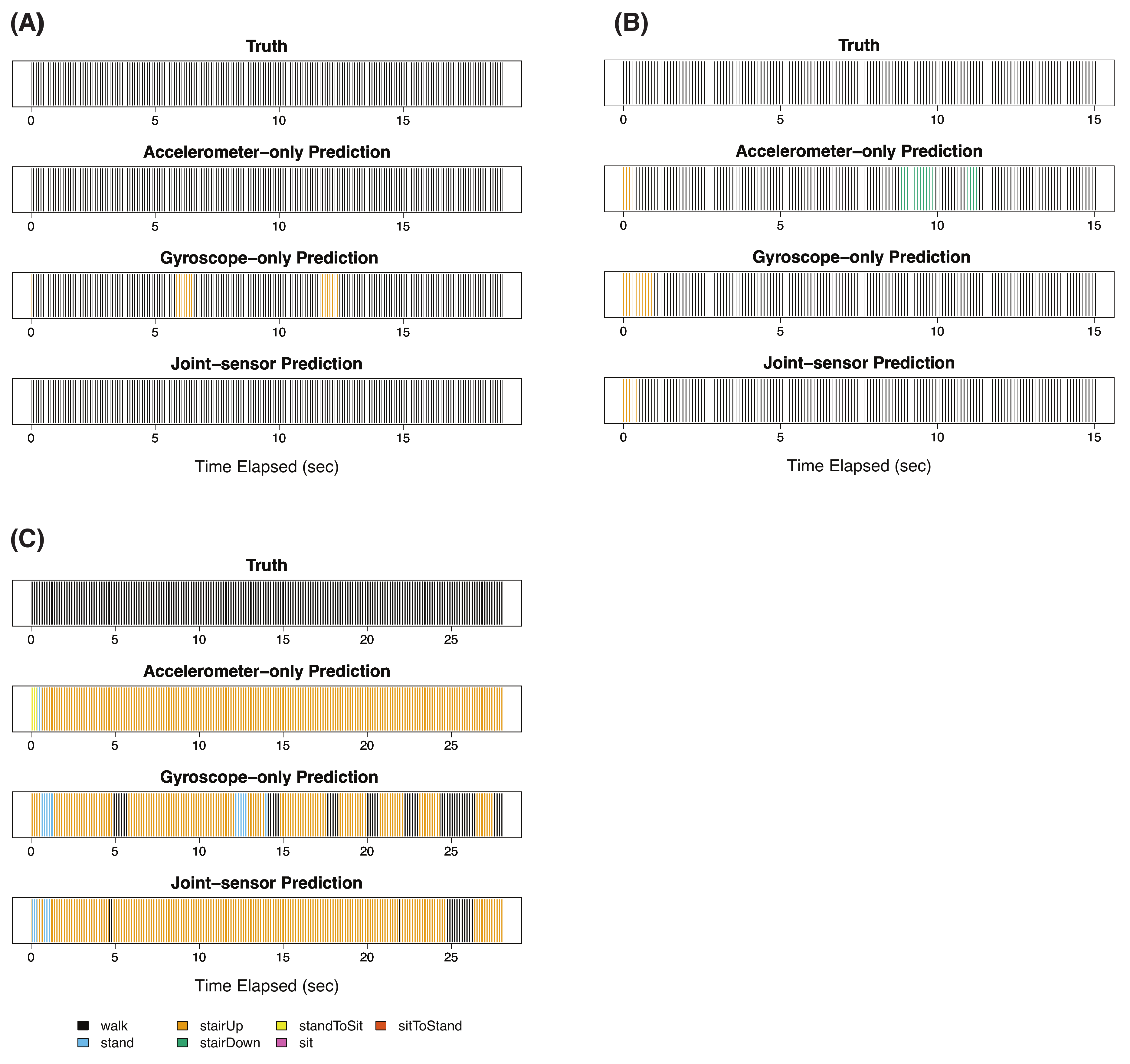}
\caption{Step 3 for Participant 3. \textit{This figure presents the results for Participant 3 during Step 3. In this step, the participant walked at three different speeds. Panel A corresponds to normal-paced walking, Panel B to fast walking, and Panel C to slow walking.}}
\label{fig:P3_S3_CompGraph}
\end{figure}

In Step 3 (Figure \ref{fig:P3_S3_CompGraph}), the participant walked at three different speeds: normal (Panel A), fast (Panel B), and slow (Panel C). All three methods correctly classified the normal walk (Figure \ref{fig:P3_S3_CompGraph}A) for most or all of the time. The gyroscope-only method had two false positives of ascending stairs, which were corrected by the joint-sensor method. For fast walking (Figure \ref{fig:P3_S3_CompGraph}B), all three methods were accurate for most of the time, with the joint-sensor method performing the best. The gyroscope-only method incorrectly predicted a 1-second episode of ascending stairs at the beginning of the fast walk. The joint-sensor method shortened this error by about one half. The accelerometer-only method also predicted an episode of ascending stairs at the beginning, as well as two episodes of descending stairs. The joint-sensor method was able to eliminate the two erroneous episodes of descending stairs, though the ascending stair error remained. All three methods showed large errors in classifying slow walking (Figure \ref{fig:P3_S3_CompGraph}C).  The accelerometer-only method misclassified the entire slow walking period, using ascending stairs for most of the time. The gyroscope-only method also chose ascending stairs  for most of the slow walking period, though it captured some episodes of walking. The joint-sensor method performed better than the accelerometer only by classifying some brief periods as walking, but it performed worse than the gyroscope alone. One might expect that the classification gets harder when the pace of walking gets slower because the slower pace can match the pace of movement of other activities, such as ascending stairs. The three methods also had difficulty in classifying slow walking for the other participants of the study. We discuss this further in Section \ref{s:discussion}.

\begin{figure}[h!]
    \centering
    \includegraphics[scale = 0.5]{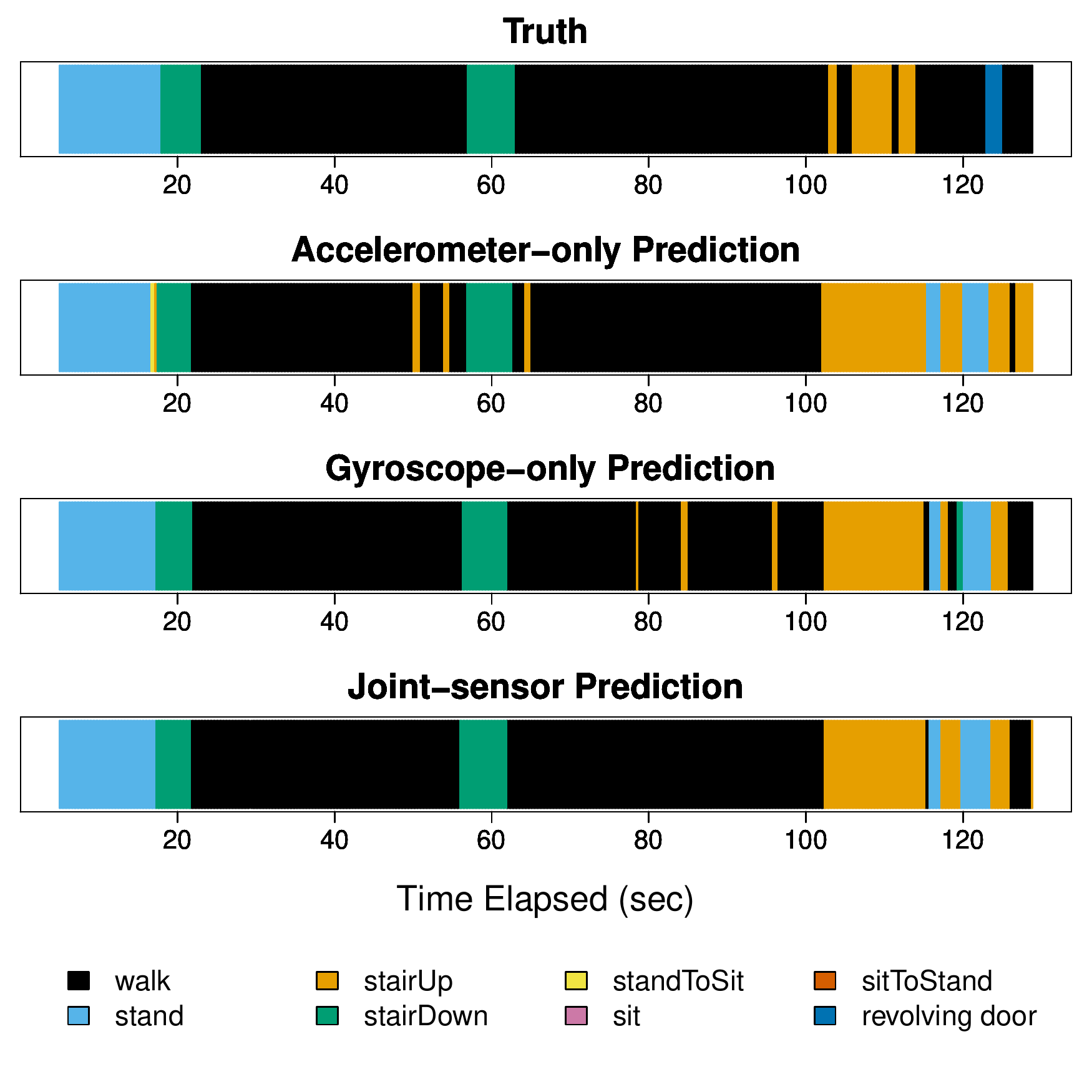}
    \caption{Step 5 for Participant 3. \textit{This figure presents the results for Participant 3 during Step 5. In Step 5, the participant performed the activities of descending stairs, walking, ascending stairs, and going through a revolving door. The revolving door activity is not included in the participant's dictionary.}}
    \label{fig:P3_S5_CompGraph}
\end{figure}

In Step 5 (Figure \ref{fig:P3_S5_CompGraph}), the participant performed the activities of standing (blue), descending stairs (green), walking (black), ascending stairs (orange), and going through a revolving door (dark blue). All three methods accurately classified the two periods of descending stairs. They also performed well at classifying the two extended walking periods, with the joint-sensor method performing the best. During these walking periods, the accelerometer-only method and gyroscope-only method each sometimes confused walking for bursts of ascending stairs. These errors were corrected by the joint-sensor method. All three methods struggled at the end of Step 5 when the participant went through the revolving door (dark blue). This is because going through a revolving door was not in the dictionary, so all methods used other activities to substitute it, including standing and ascending stairs. There was a walking (black) period directly before the participant went through the revolving door. The three methods misclassified most of this walking period using standing and ascending stairs. This may be due to the participant's reduction in walking speed before entering the revolving door.

\begin{figure}[h!]
    \centering
    \includegraphics[width=\linewidth]{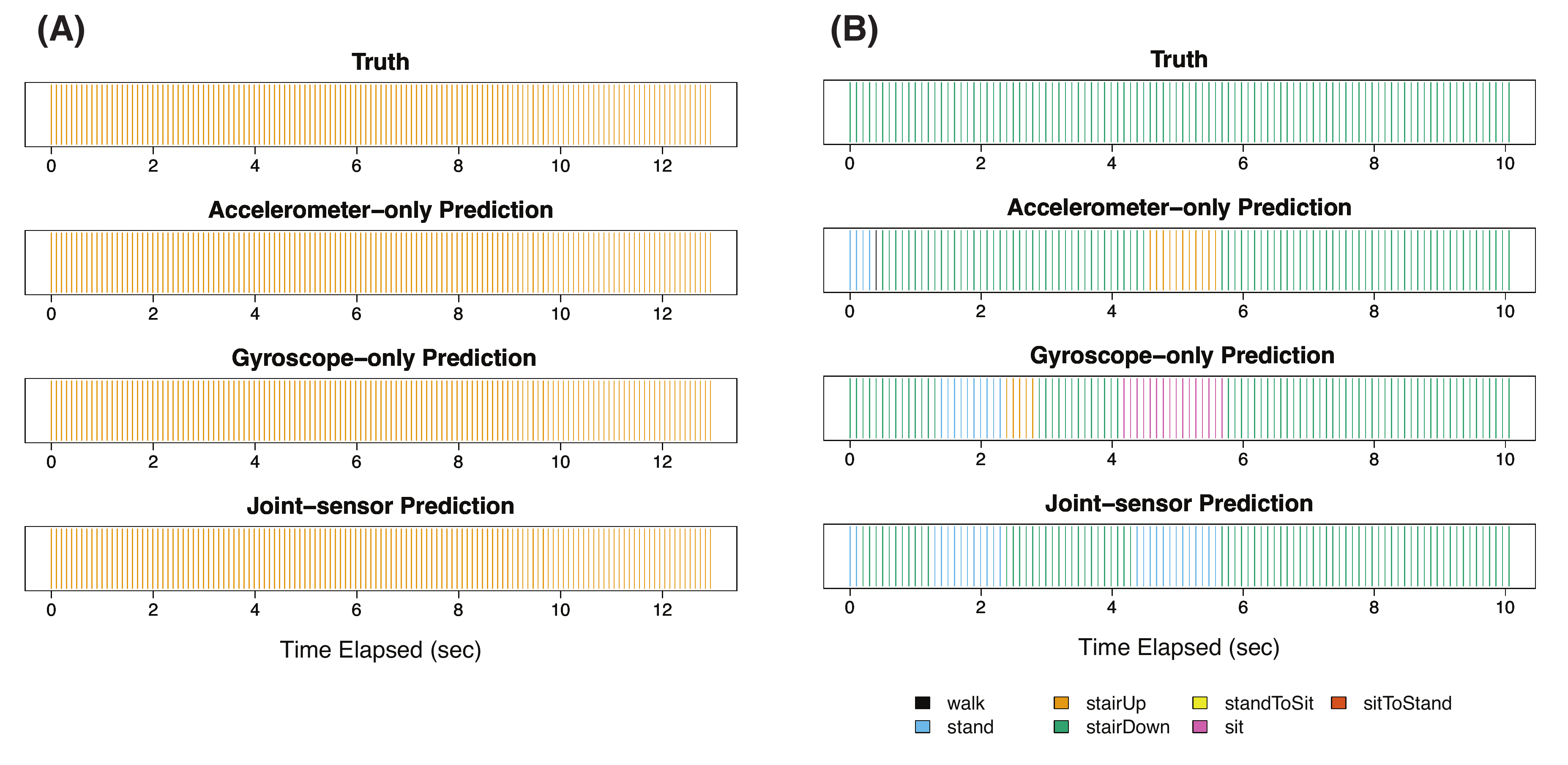}
\caption{Step 6 for Participant 3. \textit{This figure presents the results for Participant 3 during Step 6. In Step 6, the participant ascended and descended a staircase. Panel A corresponds to ascending stairs and Panel B to descending stairs.}}
\label{fig:P3_S6_CompGraph}
\end{figure}

In Step 6 (Figure \ref{fig:P3_S6_CompGraph}), the participant ascended (Panel A) and descended (Panel B) a staircase. All three methods correctly classified ascending stairs, as shown in Figure \ref{fig:P3_S6_CompGraph}A. The methods also correctly classified descending stairs (Figure \ref{fig:P3_S6_CompGraph}B) for most of the time, with the accelerometer-only performing the best. The descending stair period in Step 6 was correctly classified for about 84\% of the time using the accelerometer alone, 70\% using the gyroscope alone, and 75\% using the sensors jointly. The accelerometer-only method confused the beginning of the period with standing, and the middle portion with ascending stairs. The gyroscope-only method  had more errors, including episodes of standing, ascending stairs, and sitting.  Combining the sensors improved the results from the gyroscope alone, eliminating the ascending stair error, as well as shortening the length of the sitting error and changing it to standing, which is closer to the ground truth activity of descending stairs. Combining the sensors  reduced the length of the standing error from the accelerometer-only method. However, it also added a new standing error at about 1.3 seconds elapsed, originating from the gyroscope data.

\begin{figure}[h!]
    \centering
    \includegraphics[width=\linewidth]{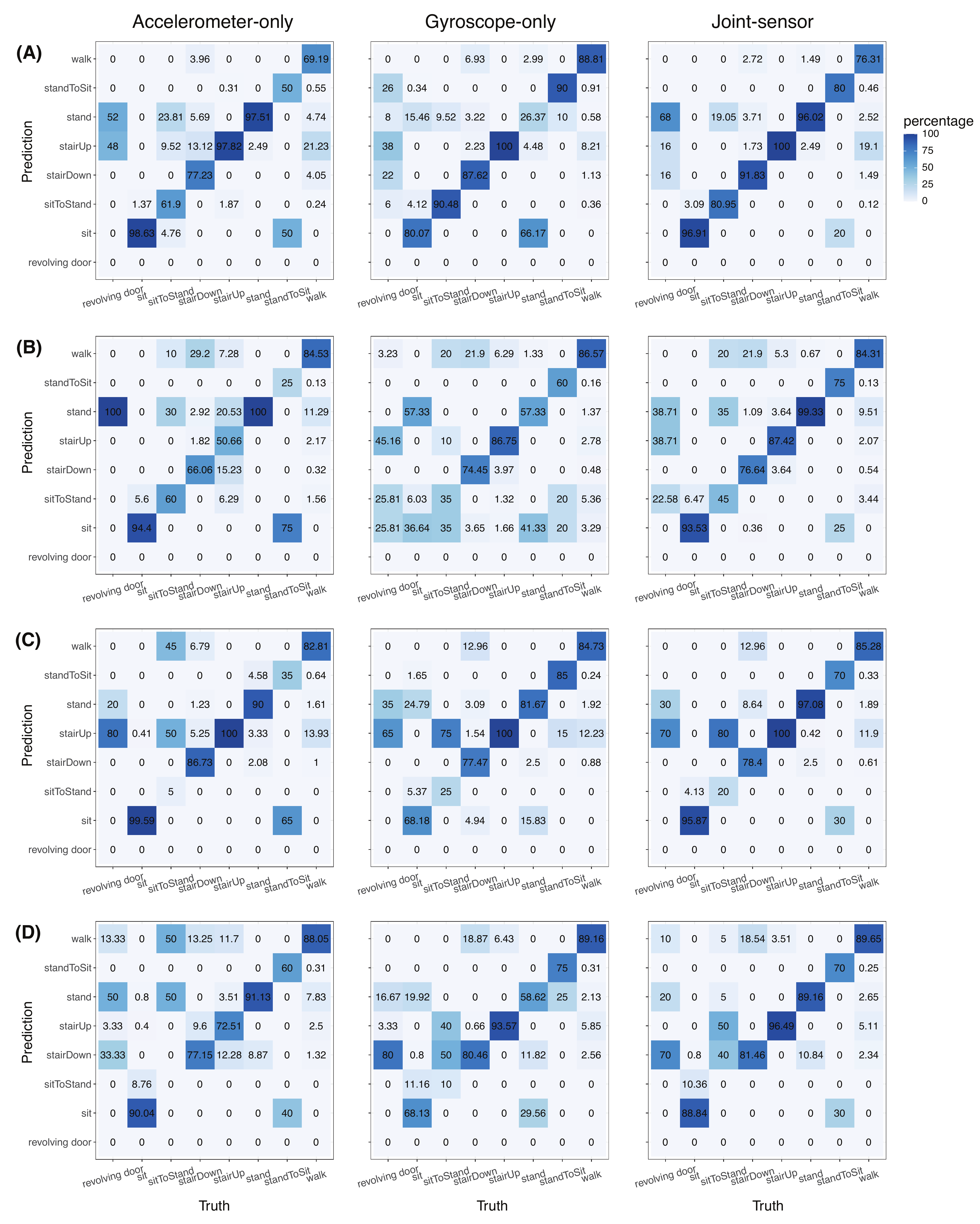}
\caption{Confusion Matrices for Accelerometer-Only, Gyroscope-Only, and Joint-Sensor Methods. \textit{This figure presents the confusion matrices for each participant. Row A corresponds to Participant 1, Row B to Participant 2, Row C to Participant 3, and Row D to Participant 4. For each participant, there are three confusion matrices corresponding to accelerometer-only (Column 1), gyroscope-only (Column 2), and joint-sensor (Column 3). These confusion matrices incorporate Steps 1, 2, 3, 5, and 6 in test data collection. In each confusion matrix, the ground truth activity labels are on the bottom margin, and the predicted activity labels are on the left margin. Each column shows the distribution of the predicted labels for the corresponding ground truth activity label. Thus, every column sums to 100\%.}}
\label{fig:allSteps_ContTable}
\end{figure}

The results of Participant 3 from Steps 1, 2, 3, 5, and 6 are quantified in row C of Figure \ref{fig:allSteps_ContTable}. The figure includes confusion matrices for the accelerometer-only (Column 1), gyroscope-only (Column 2), and joint-sensor (Column 3) methods. Each row corresponds to a different participant, with Participant 3 in row C. In each confusion matrix in Figure \ref{fig:allSteps_ContTable}, any given column corresponds to a unique ground truth activity label, and the values in the column show the distribution of the predicted activity labels for the given ground truth activity label. We measure accuracy by the diagonal elements of the confusion matrices. 

The results for Participant 3 in Figure \ref{fig:allSteps_ContTable} are consistent with our observations from Figures \ref{fig:P3_S1_S2_CompGraph} - \ref{fig:P3_S6_CompGraph}. For example, the gyroscope-only method (row C, column 2) tended to confuse standing and sitting, classifying 25\% of the ``sit" labels as ``stand" and 16\% of the ``stand" labels as ``sit". In contrast, the joint-sensor method (row C, column 3) classified standing correctly 97\% of the time and sitting correctly 96\% of the time. The accelerometer-only method (row C, column 1) had high accuracies for most of the activities (i.e., dark shading in the diagonal entries), but classified the ``sit-to-stand" labels with only 5\% accuracy. All three methods classified the ascending stair (``stairUp") labels correctly. The accelerometer-only method had higher accuracy than the gyroscope-only method for standing, sitting, and descending stairs. The gyroscope-only method had higher accuracy than the accelerometer-only method for walking, sit-to-stand, and stand-to-sit. For all activities, the accuracy value of the joint-sensor method was either between that of the accelerometer-only and gyroscope-only methods, or was higher than both of them.  

\subsection{Quantifying Classification Accuracy}
\label{s:results_allParticipants}

\begin{figure}[h!]
    \centering
    \includegraphics[scale = 0.5]{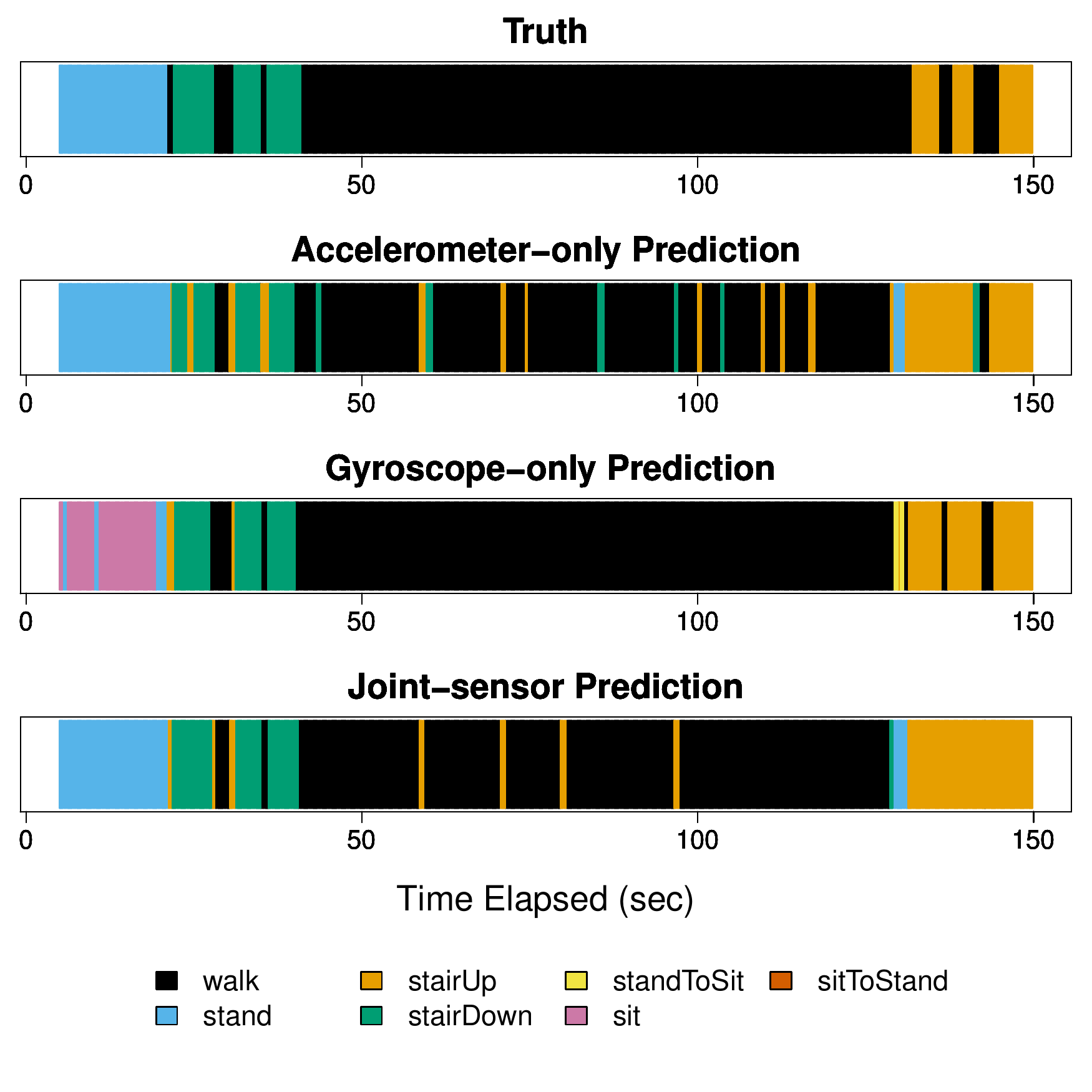}
    \caption{Step 1 for Participant 1. \textit{The figure shows the results for Participant 1 during Step 1.}}
    \label{fig:P1_S1_CompGraph}
\end{figure}

In this section, we discuss the results for the other participants. The confusion matrices for Participants 1, 2, and 4 are shown in rows A, B, and D, respectively, of Figure \ref{fig:allSteps_ContTable}. Also, the counterparts to Figures \ref{fig:P3_S1_S2_CompGraph} - \ref{fig:P3_S6_CompGraph} for these other participants are shown in the Supplement (Figures S1 - S8). There are similarities across the participants. For instance, as with Participant 3, the gyroscope-only method tended to mix up standing and sitting for all participants. The joint-sensor method was able to correct most of these errors. We also observed that the joint-sensor method could correct systematic errors from one sensor in misidentifying walking with another activity.  In Participant 3, the classifications of the gyroscope-only method sometimes misrepresented walking as short bursts of ascending stairs, as discussed in Section \ref{s:results_participant3}. The joint-sensor method was able to correct many of these errors. This same pattern of correction also occurred for Participants 1 and 4. One should note that the systematic errors were not exclusive to one particular sensor. An interesting difference between the participants was that, while the short errors mostly occurred from gyroscope-only for Participant 3, they tended to be from the accelerometer-only for Participants 1 and 4. As an example, we show the Step 1 result for Participant 1 in Figure \ref{fig:P1_S1_CompGraph}. For walking (black), the accelerometer-only method characterized many short segments as ascending stairs or descending stairs. The joint-sensor method reduced the number of these false positives that occurred in the classifications from accelerometer alone. We believe the reason that the joint-sensor method could generate improvement was that it was rare for both sensors to misidentify walking at the same time with the same activity. The error might occur in accelerometer only or gyroscope only, but not both. 

The joint-sensor method did not always achieve superior results over a single-sensor method. This was particularly the case for Participant 1.  For instance, in Figure \ref{fig:P1_S1_CompGraph} (Participant 1, Step 1), the joint-sensor method slightly degraded the walking classifications by the gyroscope only. This is because some false positives of ascending stairs originating from the accelerometer data remained in the joint-sensor classifications. We also found that some short periods of walking between ascending stairs that were picked up by the gyroscope were smoothed out by the joint-sensor method. On the other hand, the joint-sensor method still has advantages over using a single sensor. This is because which single-sensor method (accelerometer or gyroscope) performed better was not consistent across participants. In such situations, the joint-sensor method generally gave a closer result to the better performing single-sensor method for the given participant.

Table \ref{table:avgAccuracy} summarizes the results from Figure \ref{fig:allSteps_ContTable} by looking at the activities in groups. Four activity groups were considered, including (A) all activities, (B) vigorous activities, (C) stationary activities, and (D) transition activities. The table presents the average accuracy value of each method for each participant in each activity group. We first focus on Section A of Table \ref{table:avgAccuracy}, which corresponds to the all activities group. This group consists of walking, standing, ascending stairs, descending stairs, sitting, the sit-to-stand transition, and the stand-to-sit transition. Thus, the average accuracy in Section A was computed by averaging all the diagonal elements of the corresponding confusion matrix, except for ``revolving door''. The revolving door activity was excluded since it was not part of the participants' dictionaries. For the all activities group, the accelerometer-only and gyroscope-only columns are similar to each other. This means that, for each participant, the average accuracy value for the accelerometer-only method was close to that of the gyroscope-only method. The average accuracy tended to improve after combining the accelerometer and gyroscope data. The degree of improvement varied by participant. The percent increase in average accuracy from the joint-sensor method (relative to the higher of accelerometer-only and gyroscope-only) was 10.4\% for Participant 1, 16.7\% for Participant 2, 4.7\% for Participant 3, and 7.7\% for Participant 4.   

To assess which activities yielded the largest improvements, we divided the activities into smaller groups, including vigorous activities, stationary activities, and transition activities. The vigorous activities included walking, ascending stairs, and descending stairs. The stationary activities included standing and sitting. Lastly, the transition activities included stand-to-sit and sit-to-stand. We present the average accuracy values for each of these activity groups in Table \ref{table:avgAccuracy}, with vigorous activities in Section B, stationary activities in Section C, and transition activities in Section D.

\begin{table}[h]
    \centering
    \begin{tabular}{c | c | c | c | c}
    & Participant & Accelerometer & Gyroscope & Joint-sensor\\
    \hline \hline
    (A) All Activities & 1 & 78.9 & 80.5 & 88.9\\
    & 2 & 68.7 & 62.4 & 80.2\\
    & 3 & 71.3 & 74.6 & 78.1\\
    & 4 & 68.4 & 67.8 & 73.7 \\
    \hline \hline
    (B) Vigorous Activities & 1 & 81.4 & 92.1 & 89.4\\
    & 2 & 67.1 & 82.6 & 82.8\\
    & 3 & 89.8 & 87.4 & 87.9\\
    & 4 & 79.2 & 87.7 & 89.2\\
    \hline \hline
    (C) Stationary Activities  & 1 & 98.1 & 53.2 & 96.5\\
    & 2 & 97.2 & 47.0 & 96.4\\
    & 3 & 94.8 & 74.9 & 96.5\\
    & 4 & 90.6 & 63.4 & 89.0\\
    \hline \hline
    (D) Transition Activities & 1 & 56.0 & 90.2 & 80.5\\
    & 2 & 42.5 & 47.5 & 60.0\\
    & 3 & 20.0 & 55.0 & 45.0\\
    & 4 & 30.0 & 42.5 & 35.0\\
    \hline
    \end{tabular}
    \caption{Average accuracy of each method for different activity groups. \textit{The vigorous activities included walking, ascending stairs, and descending stairs. The stationary activities included sitting and standing. The transition activities included stand-to-sit and sit-to-stand.}}
    \label{table:avgAccuracy}
\end{table}

For the vigorous activities (Table \ref{table:avgAccuracy}, Section B), the joint-sensor method had higher average accuracy than the accelerometer-only method for all participants, except for Participant 3. The percent improvement in average accuracy was 9.8\% for Participant 1, 23.4\% for Participant 2, and 12.6\% for Participant 4. For Participant 3, there was a decrease in average accuracy of 2.1\%. For all participants, the average accuracy of the joint-sensor method was close to that of the gyroscope-only method. Relative to the gyroscope-only method, there was a 2.9\% decrease in average accuracy for Participant 1, a 0.24\% increase for Participant 2, a 0.57\% increase for Participant 3, and a 1.7\% increase for Participant 4. Thus, for the vigorous activities,
the improvement of the joint-sensor method was mostly from correcting the errors of the accelerometer-only method. 

For the stationary activities (Table \ref{table:avgAccuracy}, Section C), the average accuracy for the joint-sensor method was considerably higher than for the gyroscope-only method. This was because the mix-up between standing and sitting was corrected by including accelerometer data. For example, the average accuracy of Participant 2 was 97.2\% for accelerometer-only, 47.0\% for gyroscope-only, and 96.4\% for the joint-sensor method. Hence, for the stationary activities, the improvement of the joint-sensor method was mainly from correcting the errors of the gyroscope-only method. 

For the transition activities (Table \ref{table:avgAccuracy}, Section D), the average accuracy values of the joint-sensor method tended to be low for most participants. When used on their own, the sensors each had difficulty recognizing the transition activities as well. The low accuracy rates may partly be due to the movelet method. For example, although the method might detect the transition activities, the classifications may be shifted slightly too soon in time. This is because the classification at any given time point involves taking a majority vote among the 10 movelets after the time point. Since the transition activities are momentary, some movelets could be picking up the activity occurring after the transition.  

Overall, we find that the different sensors play different roles in correcting classification errors. The joint sensor method can correct the shortcomings of the gyroscope-only method in standing and sitting. It can also correct the shortcomings from the accelerometer-only method in the vigorous activities. These findings are consistent with the physical nature of the sensors.

\section{Discussion}
\label{s:discussion}

This study found that combining accelerometer and gyroscope data can result in more accurate activity recognition. For example, the gyroscope-only method has difficulty in differentiating between the activities of standing and sitting, but combining accelerometer with gyroscope data largely corrected this error. For the activity of walking, combining accelerometer and gyroscope data improved the accuracy compared to the accelerometer alone in some cases (e.g., Participant 1) and to the gyroscope alone in other cases (e.g., Participant 3). Although the single-sensor methods using accelerometer or gyroscope classified ascending and descending stairs to a certain degree, the combined method using both sensors made further improvement in some cases. Our results also show that for certain types of movement, a properly chosen single-sensor method may be adequate, e.g., accelerometer for stationary activities. These findings highlight the close connection among the specification of scientific questions (what activities are of interest), the choice of data types (whether to collect accelerometer data, gyroscope data, or both), and the choice of data analysis method.

In future work, we plan to add new extensions to the joint-sensor method. Based on our results, gyroscope and accelerometer data seemed to play different roles in identifying different types of movement. To take advantage of these differences, an area of future work is to assign different weights to the two sensors and their axes ($x$, $y$, or $z$) in the joint-sensor method. Also, we used linear interpolation to interpolate the gyroscope data to the accelerometer timestamps in this analysis. An area of future work is to consider other interpolation methods, such as cubic splines or B-splines, and conduct a sensitivity analysis about the effect of the interpolation method on the results. 

In the analysis results, the joint-sensor method and the two single-sensor methods all had difficulty with accurately classifying slow walking. To address this challenge, one approach could be to stretch or compress the 1-second dictionary movelets, which is referred to as movelet transforms \citep{karas2021}. The purpose of movelet transforms is to improve activity recognition in cases where the participant performs a given activity at a different pace during testing, compared to during training. These transforms might enable more accurate recognition of slow walking. Also, they may be useful in cases where a patient's condition is evolving over time (e.g., a patient's walking pace may increase over time as she/he recovers from surgery). 

In this paper, we studied the specific case that the phone is worn in the pocket. In reality, the phone can be carried in different locations (e.g., pocket, hand, backpack, purse) that can change with time. The specific context may also differ (e.g., phone in a tighter pocket or oriented in a different direction). An area of future work is to extend the joint-sensor method to accommodate these changes robustly. One approach is to identify the location of smartphone placement based on the accelerometer and gyroscope data, and then apply the appropriate dictionary accordingly. We may also consider standardizing the training and testing data based on the phone's placement to reduce the context influence on the amplitude.

This work was a pilot study using a relatively small sample size collected by the investigators. Our goal was to understand each sensor's role and how the combination of the sensors could provide further information. To achieve this goal, we performed a detailed analysis at the highest possible frequency, verifying the activity classification at each time point and for each activity. In future work, we will establish  the performance measures of the joint-sensor method, such as its sensitivity, specificity, and precision for each activity type. For this statistical analysis, we will apply the method to a large sample size of diverse participants that will be collected systematically through an institutional effort. With more sampled participants, we may explore the possibility of finding more general dictionaries through a clustering analysis. We may further use machine learning to automate the dictionary generation process for each individual person.

\section*{Funding}

E.J.H. and J.O. were supported by Grant U01HL145386, awarded by the National Heart, Lung, and Blood Institute (NHLBI). K.Y. was supported by the URECA Center at Wake Forest University. This paper’s contents are solely the responsibility of the authors and do not represent the views of these organizations.

\section*{Supplementary Materials}

The Supplementary Materials are available upon request (huange@wfu.edu).

\bibliographystyle{apalike}
\bibliography{references}

\end{document}